\documentclass[onecolumn,11pt]{article}
\newcommand{\ket}[1]{|#1\rangle}
\newcommand{\bq}{\begin{quote}}
\newcommand{\eq}{\end{quote}}
\newcommand{\be}{\begin{equation}}
\newcommand{\ee}{\end{equation}}
\newcommand{\ben}{\begin{enumerate}}
\newcommand{\een}{\end{enumerate}}
\newcommand{\braket}[2]{\langle#1|#2\rangle}
\title{Quantum Mechanics and Experience}
\author{Ulrich Mohrhoff\\
\textit{\small Sri Aurobindo International Centre of Education}\\ 
\textit{\small Pondicherry 605002 India}\\
\ttfamily{\small ujm@auromail.net}}
\date{}
\normalsize
\begin{document}
\maketitle
\begin{abstract}
Whether we want to make sense of the presence of consciousness in a seemingly material world or understand the role (if any) that consciousness plays in the fundamental theoretical framework of contemporary physics, it is imperative that we distinguish between two concepts of reality: an epistemically inaccessible transcendental reality and an empirical reality experienced and objectified by us. After a summary of Bohr's views and their relation to Kant's theory of science, two fruitless lines of attack on the measurement problem are discussed: the way of the $\psi$-ontologist and the way of the QBist. In the remainder of the paper the following results are obtained. (i) Because the testable correlations between outcomes of measurements of macroscopic positions are consistent with both the classical and the quantum laws, there is no conflict between the superposition principle and the existence of measurement outcomes. (ii) Intrinsically, each fundamental particle is numerically identical with every other fundamental particle. What presents itself here and now with these properties and what presents itself there and then with those properties is one and the same entity, herein called ``Being.'' (iii)  The distinction between a classical domain and a quantum domain is essentially a distinction between the manifested world and its manifestation. By entering into reflexive spatial relations, Being gives rise to (a) what looks like a multiplicity of relata if the reflexive quality of the relations is ignored, and (b) what looks like a substantial expanse if the spatial quality of the relations is reified. (iv) The reason why quantum mechanics is a calculus of correlations between measurement outcomes is that it concerns the progressive realization of distinguishable objects and distinguishable regions of space. (v) The key to the relation between quantum mechanics and experience is that Being does not simply manifest the world; Being manifests the world to itself. It is at once the single substance by which the world exists and the ultimate self or subject for which it exists. The question how we are related to this ultimate self or subject is discussed.
\end{abstract}

\section{Introduction}
Claims about the relation between quantum mechanics and experience are generally made in either of two contexts. The first context is the interpretation of quantum mechanics. ``The dynamics by itself'' \cite{Albert} does not seem to yield the definite measurement outcomes that we experience, and our experience does not seem to be compatible with the superpositions that quantum mechanics predicts. Quantum mechanics appears to require preternatural interventions into the course of Nature, and the obvious perpetrator of such interventions is the consciousness of the observer. Today the most prominent exponent of this relation between quantum mechanics and experience is Stapp~\cite{Stapp}. The second context is the problem of explaining how objective physical processes in a brain give rise to subjective experiences. In this context it has been argued (to give just one example) that quantum coherence in microtubules might be the neural basis for consciousness~\cite{Hameroff}. In both contexts it is assumed that the aim of science is to describe a mind-independent reality, and that in addition there is consciousness or experience. In the first context consciousness is invoked to solve the notorious measurement problem, while in the second context it is invoked to explain the mysterious presence of consciousness in a seemingly material world. 

In this paper I argue that it is \emph{not} the aim of science to describe a mind-independent reality. Rather, science is concerned with the objectifiable aspects of our experience. Whether we want to make sense of the presence of consciousness in a seemingly material world or to grasp the role (if any) that experience plays in the fundamental theoretical framework of contemporary physics, it is imperative that we distinguish between two concepts of reality: the objectified world-as-we-know-it and the epistemically inaccessible world-in-itself. The necessity of drawing this distinction was first emphasized by Immanuel Kant in the 18th Century and was reaffirmed by Niels Bohr in the 20th. Section \ref{sec_Bohr} contains a brief summary of Bohr's views, and Sect.\ \ref{sec_Kant} discusses the relation between Bohr's views and Kant's theory of science.

Failure to distinguish the middle-ground between the experiencing and objectifying subject and the world-in-itself---i.e., failure to acknowledge that physical science concerns a reality experienced and objectified by us---gives rise to two fruitless lines of attack on the quantum measurement problem: the way of the ``$\psi$-ontologist'' and the way of the QBist or Quantum Bayesian. The former looks upon the evolution of the wave function as a physical process taking place in a mind-independent reality, the latter looks upon quantum theory as an addition to Bayesian probability theory. After a critique of the first line of attack, Sect.\ \ref{sec_mmproblem} counters a QBist critique of Bohr, and Sect.\ \ref{sec_QBism} offers a critique of the salient points of QBism.

If the preconceptions that go into the interpretation of a theory themselves obey the laws of the theory, the theory is said to be semantically consistent. If quantum mechanics is to be semantically consistent then even outcome-indicating observables must obey the theory's correlation laws, in spite of the intrinsic definiteness of our experience, which would seem to imply that such observables are exempt from the superposition principle. In Sect.\ \ref{Sec_sc} the semantic consistency of quantum mechanics is demonstrated by first showing that the theory implies the incompleteness of the world's spatiotemporal differentiation: this does not go ``all the way down.'' Hence follows the existence of macroscopic objects, defined as objects whose trajectories are only counterfactually indefinite: no value-indicating event reveals the indefiniteness of a macroscopic position in the only way it could, through a departure from what the classical laws predict. The testable correlations between outcomes of measurements of macroscopic positions are therefore consistent with \emph{both} the classical and the quantum laws, so that there is no conflict between the superposition principle and the existence of measurement outcomes.

We are now in a position to think about the objectifiable implications of the quantum-mechanical correlation laws without having to keep in mind our role as experiencing and objectifying subjects. The safest way to proceed is to use that formulation of quantum mechanics which carries the least metaphysical baggage, which is due to Feynman. We also need an interpretive principle, which is introduced in Sect.\ \ref{sec_qr} and applied to two paradigmatic set\-ups, one concerning distinctions between {regions of space}, the other concerning distinctions between {things}. The main conclusion of this section is that, intrinsically, \emph{each fundamental particle is numerically identical with every other fundamental particle}. What presents itself here and now with these properties and what presents itself there and then with those properties is one and the same entity, herein called ``Being.''

The main conclusion of Sect.\ \ref{sec_manif} is that the distinction between the macroscopic or classical domain and the microscopic or quantum domain is essentially a distinction between the manifested world and its \emph{manifestation}. It is by entering into reflexive spatial relations---i.e., self-relations, relations between numerically identical relata---that Being gives rise to (i)~what looks like a multiplicity of relata if the reflexive quality of the relations is ignored, and (ii)~what looks like a substantial expanse if the spatial quality of the relations is reified. Seen in this light, quantum theory reverses the explanatory arrow of classical or folk physics: instead of explaining wholes in terms of interacting parts, it suggests that the multiplicity of the world emerges from an intrinsically undifferentiated reality, through a series of stages in which the world's differentiation into distinguishable regions of space and distinguishable objects with definite properties is progressively realized. \looseness=1 

We next turn to the question what quantum mechanics may \emph{actually} have to do with experience. The most promising alternative to reductionism in the philosophy of mind is panpsychism. Noting that all physical properties are relational or extrinsic, proponents of panpsychism embrace the possibility of situating consciousness among the intrinsic properties of the relata which bear the relational properties. The problem with this possibility is the difficulty of explaining how the consciousnesses of a myriad of particles can be combined into a complex and rich consciousness such as that we possess \cite{Seager}. If however we take into account not only that all physical properties are relational but also that all relational properties are reflexive, so that the relata are identically the same Being, the concept of  consciousness as an intrinsic property of the relata comes into its own. 

How does this intrinsic property of Being relate to the manifestation? The answer is that Being does not simply manifest the world; Being manifests the world \emph{to itself}. Being relates to the world not only as the substance that constitutes it but also as the consciousness that contains it. It is at once the single substance by which the world exists and the ultimate self or subject for which it exists. The question then is how we, as conscious beings, are related to this ultimate self or subject, and this is addressed in Sect.\ \ref{sec_exp}, which takes its cue from the Indian philosopher (and freedom fighter, and mystic) Sri Aurobindo \cite{Heehs2008, SA_LD}. 

The concluding section recapitulates arguments to the effect that the validity of quantum mechanics is but a necessary consequence of the manner in which---and the purpose for which---the physical world came into being \cite{Mohrhoff-justso, Mohrhoff-QMexplained, Mohrhoff-book}. Finally, our attempts to make sense of the world in which we find ourselves are put in perspective.

\section{Bohr and ``experience''}\label{sec_Bohr}
\bq
\emph{Bohr's claim was that the \emph{classical language is indispensable}. This has remained valid up to the present day. At the \emph{individual} level of clicks in particle detectors and particle tracks on photographs, all measurement results have to be expressed in classical terms. Indeed, the use of the familiar physical quantities of length, time, mass, and momentum--energy at a subatomic scale is due to an extrapolation of the language of classical physics to the non-classical domain.}\hfill--- Brigitte Falkenburg~\cite[p.\ 162]{Falk}
\eq
Kant taught us to distinguish between two concepts of reality, a transcendental reality and an empirical one, and Bohr stressed the need for this distinction when he reminded us that ``in our description of nature the purpose is not to disclose the real essence of the phenomena but only to track down, so far as it is possible, relations between the manifold aspects of our experience'' \cite[pp.\ 17--18]{Bohr-ATDN}. Bohr's reminder has  been misconstrued nearly as often as it has been quoted. A recent example is Mermin~\cite{Mermin_Nature}, who in a comment published in \emph{Nature} maintains that Bohr ``taught that physical science studies our experience.'' What would have been more correct to say is that physical science, according to Bohr, studies not our experience \emph{qua experience} but the \emph{objectifiable aspects} of our experience. The reason Bohr refers to the relations between the manifold aspects of our experience is that these relations make it possible to objectify certain aspects of our experience---to represent them as aspects of a reality from which we, the objectifying subjects, can abstract ourselves. Because of our familiarity with the logical/grammatical relation between a subject and its predicates it makes sense to us to think of the content of our experience as the properties of \emph{substances}. Because of our familiarity with the conditional relation between propositions it makes sense to us to think of the successive (or timelike related) aspects of our experience as properties related by \emph{causality}, and it makes sense to us to think of the simultaneous (or spacelike related) aspects of our experience as a self-existent system of \emph{interacting} substances. It is because Bohr considered the categorial structure implicit in the language of classical physics a \emph{sine qua non} of empirical science that he insisted on the use of this language in describing experiments and reporting their results. This structure is in all significant respects identical with the system of categories that Kant had shown to be a precondition of the possibility of empirical knowledge.%
\footnote{In a talk given three months after the publication of his \emph{Nature} comment, Mermin~\cite{Mermin_QBism} takes a stab at elucidating Bohr's meaning of ``experience'': ``by `experience' he almost certainly meant the objective readings of large classical instruments.'' Taken together, the two statements---that physical science studies our experience, and that experience stands for the objective readings of large classical instruments---imply that, according to Bohr, physical science studies the objective readings of large classical instruments, which is a crude statement of instrumentalism and an insult to Bohr.}

To determine the import of Bohr's reminder, one has to take account of its context~\cite[p.\ 17--18]{Bohr-ATDN}:
\bq
Now, what gives to the quantum-theoretical description its peculiar characteristic is just this, that in order to evade the quantum of action we must use separate experimental arrangements to obtain accurate measurements of the different quantities, the simultaneous knowledge of which would be required for a complete description based upon the classical theories, and, further, that these experimental results cannot be supplemented by repeated measurements. \dots a subsequent measurement to a certain degree deprives the information given by a previous measurement of its significance for predicting the future course of the phenomena. Obviously, these facts not only set a limit to the \emph{extent} of the information obtainable by measurements, but they also set a limit to the \emph{meaning} which we may attribute to such information. We meet here in a new light the old truth that in our description of nature the purpose is not to disclose the real essence of the phenomena but only to track down, so far as it is possible, relations between the manifold aspects of our experience.
\eq
What occupied Bohr was ``the task of bringing order into an entirely new field of experience'' \cite[p.\ 228]{Schilpp}, a field of experience that places limits on ``the \emph{extent} of the information obtainable by measurements'' and on ``the \emph{meaning} which we may attribute to such information.'' When Bohr spoke of a ``field of experience,'' when he insisted that it ``lies in the nature of physical observation, that all experience must ultimately be expressed in terms of classical concepts''~\cite{Bohr-ATDN}, and when he pointed out that ``the requirement of communicability of the circumstances and results of experiments implies that we can speak of well defined experiences only within the framework of ordinary concepts''~\cite{Bohr1937}, he was concerned with the objectifiable content of our experience, i.e., an objective reality inevitably conditioned by our experience and the concepts at our disposal but capable of being thought and spoken about without explicit reference to its being experienced or thought and spoken about.

The key to making sense of Bohr, and quite possibly the key to making sense of quantum mechanics itself, is the distinction between two concepts of reality:
\begin{itemize}
\item[(I)] a \emph{transcendental} reality undisclosed in experience, which forms or includes the ground of our experience and of ourselves as experiencing subjects, and
\item[(II)]the product of a mental synthesis, based on the spatiotemporal structure of our experience, achieved with the help of relational concepts inherent in our language, and resulting in a knowable \emph{empirical} reality from which the objectifying subject can abstract itself.
\end{itemize}
The necessity of drawing a clear distinction between a transcendental or ``veiled'' reality and a knowable or ``empirical'' reality has more recently been stressed by d'Espagnat \cite{BdE1983, BdE1989, BdE1995, BdE2006}.

Nothing is more central to Bohr's views than the dual necessity (i)~of defining observables in terms of the experimental arrangements by which they are measured and (ii)~of describing these arrangements in classical terms. Because the limits placed on simultaneous measurements of incompatible observables imply ``a complementarity of the possibilities of definition'' \cite[p.\ 78]{Bohr-ATDN}, observables can no longer be defined without reference to the ``agency of observation'' \cite[pp.\ 54, 67]{Bohr-ATDN} (i.e., the experimental apparatus), so that the ``procedure of measurement has an essential influence on the conditions on which the very definition of the physical quantities in question rests''~\cite{Bohr1935a}. The ``distinction between object and agency of measurement'' is therefore ``inherent in our very idea of observation'' \cite[ p.\ 68]{Bohr-ATDN}. And since ``[o]nly with the help of classical ideas is it possible to ascribe an unambiguous meaning to the results of observation'' \cite[p.\ 94]{Bohr-ATDN}, ``it would be a misconception to believe that the difficulties of the atomic theory may be evaded by eventually replacing the concepts of classical physics by new conceptual forms'' \cite[p.\ 16]{Bohr-ATDN}. For summaries of Bohr's mature views, which ``remained more or less stable at least over the latter thirty years of Bohr's life'' \cite[p.\ 138]{Hooker1972}, see Kaiser~\cite{Kaiser} and MacKinnon~\cite{MacKinnon}.

\section{Bohr and Kant}\label{sec_Kant}
\bq
\emph{In my opinion, those who really want to understand contemporary physics---i.e., not only to apply physics in practice but also to make it transparent---will find it useful, even indispensable at a certain stage, to think through Kant's theory of science.}\hfill\par
{\parskip=0pt \hfill--- Carl Friedrich von Weizs\"acker \cite[p.\ 328]{vW}}
\eq
Affinities between Bohr and Kant have been noted by a number of commentators \cite{Kaiser, MacKinnon, Honner, FalkCAPR, Brock, Bitbol2010, Chevalley, Faye2008, Folse, Hooker1994}, including Cuffaro \cite{Cuffaro}, who holds that any interpretation of Bohr should start with Kant. Kant owes his fame in large part to his successful navigation between the Scylla of transcendental realism (which eventually runs up against the impossible task of explaining the subject in terms of the object) and the Charybdis of a Berkeley-style idealism. What allowed him to steer clear of both horns of the dilemma was a dramatic change of strategy. Instead of trying to formulate a metaphysical picture of the world consistent with Newton's theory, as he had done during the pre-critical period of his philosophy, he inquired into the cognitive conditions that are (i)~necessary for the possibility of empirical knowledge and (ii)~sufficient for the objective status of Newtonian mechanics. He recognized, as Bitbol~\cite{Bitbol2010} has put it, ``that the land sought by metaphysics can be foreign to us not due to the excessive distance we have with it, but rather due to its excessive proximity to us.''

Kant's reading of classical mechanics dispelled many qualms that had been shared by thinkers at the end of the eighteenth century---qualms about the purely mathematical nature of Newtonian mechanics, about its partial lack of intelligibility (e.g., action at a distance), and about such unfamiliar features as Galileo's principle of relativity. The mathematical nature of Newton's theory was justified not by the Neo-Platonic belief that the book of nature was written in mathematical language but by being a precondition of empirical knowledge. What permits us to look upon the manifold aspects of our experience as the constituents of an objective world is the mathematical regularities that exist between them. They make it possible to synthesize these aspects into a system of interacting re-identifiable objects from which the experiencing subject can remove itself. 

Similarly, Newton's refusal to explain gravitational action at a distance was due not to any incompleteness of his theory but to the fact that the only causality available to us consists in regular mathematical relations between successive states or between objects in different locations. For A to be causally related to B is for A to stand in a regular mathematical relation to B. We thus lack the concepts to \emph{explain} action at a distance. As for Galileo's principle of relativity, which asserts that motion is always relative, it is a direct consequence of the fact that regular mathematical relations only exists between particular aspects of experience, i.e., between successive events and between objects in different locations, but never between a particular aspect of experience and a general form of experience, time or space.

In the course of the nineteenth century, most philosophers and many of the scientists who cared to reflect on the nature of science came to  adopt (or, in some cases, reinvent) Kant's epistemology. It alleviated the loss of transcendental realism by justifying the use of a universal objectivist language, which made it possible to think and behave \emph{as if} transcendental realism were true. The advent of quantum mechanics dealt a severe blow to this comfortable attitude. Atoms were found to respect neither Kant's apriorism (e.g., the assertion that the possibility of empirical knowledge requires the universal validity of the law of causality) nor his principle of thoroughgoing determination, which asserts that ``among all possible predicates of things, insofar as they are compared with their opposites, one must apply to [each thing] as to its possibility'' \cite[p.\ 553]{KantCPR}. Neither Kant's apriorism nor his principle of thoroughgoing determination, however, though significant elements of his thinking, are central to his theory of science. There are therefore more ways to respond to this blow than relapsing into transcendental realism, embracing a Berkeley-style idealism, or resigning oneself to a metaphysically sterile instrumentalism. In Bohr's case, MacKinnon writes, the crisis
\bq
precipitated something akin to a Gestalt shift, a shift in focal attention from the objects studied to the conceptual system used to study them. This effect was somewhat similar to the epoch\'e advocated by Husserl. He stressed the need for philosophers to bracket the natural standpoint and consider it as a representational system, rather than the reality represented. Similarly, Bohr's realization of the essential failure of the pictures in space and time on which the description of natural phenomena have hitherto been based shifted the focus of attention from the phenomena represented to the system that served as a vehicle of representation. \cite[p.\ 97]{MacKinnon}
\eq
Bohr's ``epoch\'e'' mirrors the seminal change of strategy that had enabled Kant to navigate between the Scylla of realism and the Charybdis of idealism. Where Kant had famously stressed that ``[t]houghts without content are empty, intuitions without concepts are blind'' \cite[p.~193]{KantCPR}, Bohr could have said that without measurements the formal apparatus of quantum mechanics is empty, while measurements without the formal apparatus of quantum mechanics are blind. To look beyond the directly experienced world, one needs the formal apparatus, and one needs experimental arrangements situated in that world, ``the specification of which is imperative for any well-defined application of the quantum-mechanical formalism'' \cite[p.\ 57]{Bohr-APHK}. Instead of preventing us from looking beyond the classically describable world, the mutual exclusion of experimental procedures opens a window on what lies beyond: ``it is only the mutual exclusion of any two experimental procedures, permitting the unambiguous definition of complementary physical quantities, which provides room for new physical laws, the co-existence of which might at first sight appear irreconcilable with the basic principles of science'' \cite[p.\ 61]{Bohr-APHK}. 

For Bohr, as for Kant, there is an intimate connection between perception, conditioned by the spatiotemporal aspects of experience, and conception. Kant saw conception as organizing the content of experience into an effectively subject-independent world of re-identifiable objects and causally connected events. Bohr saw conception as providing the unambiguous language necessary for the clear-cut separation of the subject from the object:  ``all subjectivity is avoided by proper attention to the circumstances required for the well-defined use of elementary physical concepts'' \cite[pp.\ 7]{Bohr-EAPHK} Far from invalidating Kant's epistemological stance, quantum theory reminds us of the interplay of perception and conception in the construction of the objective world. Bohr's departure from Kant is not a departure from Kant's epistemological stance but a consequence of quantum theory's departure from classical physics. It is an irony that Bohr, seeing Kant as arguing for the a priori necessity of Newtonian physics, regarded complementarity as an alternative to Kant's theory of science, thus drawing the battle lines in a way which put Kant and himself on opposing sides.

What allowed Bohr to go beyond Kant was the discovery that empirical knowledge is not limited to what is accessible to our senses, and that it does not have to be a knowledge of interacting objects and causally connected events. What is not directly accessible to our senses, however, cannot be expected to conform to the spatiotemporal conditions of experience and to the concepts depending on them, such as position and momentum, time and energy, causality and interaction. ``It is my personal opinion,'' Bohr wrote to the philosopher Harald H\o ffding ~\cite{BohrLetter}, ``that these difficulties are of such a nature that they hardly allow us to hope that we shall be able, within the world of the atom, to carry through a description in space and time that corresponds to our ordinary sensory perceptions.''

I cannot but agree with Folse when he writes~\cite{Folse}: ``It is often said that a work of genius resists categorization. If so, Bohr's philosophical viewpoint easily passes this criterion of greatness. Surely this is one of the reasons for the commonplace complaints over Bohr's `obscurity'.'' Historically, Bohr's reply \cite{Bohr1935b} to the EPR paper was taken as a definitive refutation by the physics community. By the time interpreting quantum mechanics became a growth industry, Bohr's perspective was lost. His epistemological reflections generally came to be treated on a par with a proliferating multitude of ontological interpretations of a mathematically formulated theory, and as such they could not but seem amateurish, outdated, ad hoc, bizarre, or downright irrational. This is how Jeffrey Bub \cite[pp.\ 45--46]{Bub1974} came to conclude that  \looseness=1
\bq
The careful phraseology of complementarity \dots\ endows an unacceptable theory of measurement with mystery and apparent profundity, where clarity would reveal an unsolved problem,
\eq
how Abner Shimony \cite{Shimony} came to confess that 
\bq
after 25 years of attentive---and even reverent---reading of Bohr, I have not found a consistent and comprehensive framework for the interpretation of quantum mechanics,
\eq
and how Edwin Jaynes \cite{Jaynes} came to see in ``our present QM formalism''
\bq
a peculiar mixture describing in part realities of Nature, in part incomplete human information about Nature---all scrambled up by Heisenberg and Bohr into an omelette that nobody has seen how to unscramble.
\eq
It would therefore seem appropriate to close this section with the following words by Hooker~\cite[pp.\ 132--133]{Hooker1972}:
\bq
it is a remarkable commentary on the state of confusion and misunderstanding now existing in the field that Bohr's unique views are almost universally either overlooked completely or distorted beyond recognition---this by philosophers of science and scientists alike. Despite the fact that there are almost as many philosophies of quantum theory as there are major quantum theorists, the illusion somehow persists that they are all talking about the same thing and in essentially the same way. \dots so many people can apparently read Bohr and not grasp the significance of what he was driving at.
\eq

\section{The quantum measurement problem}\label{sec_mmproblem}
\bq
\emph{The most satisfying way to end a philosophical dispute is to find a false presupposition that underlies all the puzzles it involves.}\par
{\parskip=0pt\hfill--- Bas van Fraassen \cite[p. 434]{BCVF}\par}
\eq
What distinguishes Bohr's unique views from other quantum philosophies is his recognition of an objective middle-ground between the experiencing and objectifying subject and the epistemically inaccessible object. Failure to distinguish this middle-ground---the empirical (type-II) reality---from the transcendental (type-I) reality gives rise to two fruitless lines of attack on the quantum measurement problem. The first is the way of the $\psi$-ontologist,%
\footnote{Not to be confused with a Scientologist. For the origin of this subtly suggestive term see Note~3 of Ref.~\cite{Fuchs_Perimeter}.} 
who believes the evolution of the wave function to be a physical process taking place in a mind-independent reality. That this line of attack is doomed to fail follows from, among other things, the insolubility proofs of the objectification problem due to Mittelstaedt \cite[Sect.~4.3b]{Mittelstaedt} and Busch \emph{et al.}\ \cite[Sect.~III.6.2]{BLM}. 

The peculiar meaning of ``objectification'' in the literature on the quantum measurement problem must not be confused with the regular sense of the word, which refers to the representation of a concept as a real or concrete object. The objectification problem is an artifact of the manner in which the measurement problem has been discussed ever since its first rigorous formulations in the monographs of von Neumann~\cite{vonNeumann} and Pauli~\cite{Pauli1933}. One assumes that there is such a thing as a measurement process, and that this takes place in three steps: the system or state preparation, a continuous dynamical process called ``premeasurement''  ($p$), and the appearance of an outcome called ``objectification'' ($o$):
$$
\sum_{k}c_k\ket{A_0}\ket{q_k}\stackrel{(p)}{\longrightarrow}
\sum_{k}c_k\ket{A_k}\ket{q_k}\stackrel{(o)}{\longrightarrow}
\ket{A(q)}\ket{q}.
$$
Every interpretation of quantum mechanics that attaches to a quantum state a significance beyond that of a probability algorithm, founders over the transition from the state that is supposed to complete the premeasurement to the state that is supposed to complete the measurement, in which $q$ (one of the possible outcomes $q_k$) is indicated. One of the reasons it founders is that the state vector's or wave function's dependence on time is not the continuous time-dependence of an evolving physical state. The time it depends on is the time of the measurement to the possible outcomes of which it serves to assign probabilities. As Peres~\cite{Peres84} has stressed, ``there is no interpolating wave function giving the `state of the system' between measurements''.%
\footnote{Here Peres echoes Bohr's insistence that what happens between the preparation of a system and a measurement is a holistic phenomenon, which cannot be decomposed into the unitary evolution of a quantum state and a subsequent ``collapse'' of the same: ``all unambiguous interpretation of the quantum mechanical formalism involves the fixation of the external conditions, defining the initial state of the atomic system concerned and the character of the possible predictions as regards subsequent observable properties of that system. Any measurement in quantum theory can in fact only refer either to a fixation of the initial state or to the test of such predictions, and it is first the combination of measurements of both kinds which constitutes a well-defined phenomenon.'' \cite{Bohr_1939}}
Nothing good can come out of the transmogrification of a probability algorithm into an evolving physical state. 

During a successful measurement, the apparatus changes from a neutral state, in which it is ready to perform its function, to a state in which it indicates an outcome, but this change does not follow the unitary ``evolution'' of the quantum state of a composite system into a superposition of the form $\sum_k c_k\ket{a_k}\ket{b_k}$, where $\ket{a_k}$ and $\ket{b_k}$ are eigenstates of observables $A$ and~$B$. This superposition ``obtains'' at a time $t$ only in the conditional sense that a joint measurement of $A$ and $B$, made at the time~$t$, would (or will) indicate the pair of outcomes $a_k$ and $b_k$ with probability $|c_k|^2$, and that it would (or will) indicate the pair of outcomes $a_i$ and $b_k$ with probability~0 if $i\neq k$. The strict correlation between the possible outcomes of the two observables then warrants another conditional statement, to wit: if (say) a measurement of $A$ yields $a_k$, then a measurement of $B$ would (or will) yield $b_k$ with probability~1. What it does not warrant is the claim that $B$ then \emph{has} the value~$b_k$. Probability 1 is not sufficient for ``is'' or ``has.'' The so-called eigenvalue-eigenstate link, a commonly accepted interpretive principle, must therefore be rejected.%
\footnote{Here is how this principle was formulated by Dirac~\cite[pp. 46--47]{Dirac}: ``The expression that an observable `has a particular value' for a particular state is permissible \dots\ in the special case when a measurement of the observable is certain to lead to the particular value, so that the state is an eigenstate of the observable.'' At bottom, to solve the objectification problem is to find a dynamical explanation for the transition from probability~1 to ``is'' or ``has.'' The reason there is no solution is that there can be no dynamical explanation for an interpretive principle like the eigenvalue-eigenstate link.}

The second fruitless line of attack is the way of the QBist or Quantum Bayesian \cite{Fuchs_Perimeter, CFS2002, Fuchs_Schack2013, MohrhoffQB}. QBists hold that quantum theory is an addition to Bayesian probability theory. It provides each ``user of science'' with a calculus for gambling on her respective experiences. Because the odds assigned by different users are based on their respective prior experiences, which generally differ, QBists believe that there are potentially as many quantum states for a given system as there are users of the theory. While they agree with Bohr that the events to which quantum mechanics assigns probabilities are measurement outcomes, they define ``measurement'' as an ``action an agent takes to elicit a set of possible experiences,'' and they define a ``measurement outcome'' as ``the particular experience of that agent elicited in this way''~\cite{FMS}. Because measurement outcomes are thus relative to individual users/agents, a user/agent has to subject everything but her ``own direct internal awareness of her own private experience'' to the superposition principle, including reports of measurement outcomes she has yet to receive from other users/agents. 

QBists are wrong on several points, to be discussed in the next section, and in their critique of Bohr (``Copenhagen''), to be discussed in the remainder of this section. QBists blame the trouble we face in our attempts to make sense of quantum mechanics on ``our ingrained practice of divorcing the objects of our investigations from the subjective experiences they induce in us''~\cite{FMS}. In doing so, they conflate the type-II reality investigated by us with the type-I reality inducing experiences in us.%
\footnote{This conflation is likewise responsible for the insolubility of the ``hard problem of consciousness,'' which most philosophers of mind take to be the problem of explaining how objective physical processes in a brain give rise to subjective experiences. Whatever gives rise to experiences belongs to the epistemically inaccessible (type-I) reality, whereas the processes investigated by science belong to the empirical (type-II) reality, which being distilled from our experiences cannot give rise to experiences.}
What is responsible for our interpretive predicament is not the practice of divorcing the objects of our investigations from our subjective experiences but the practice of \emph{not} distinguishing between the objects of our investigations, which form part of the type-II reality constructed by us, and whatever it is that induces subjective experiences in us, which is or belongs to the epistemically inaccessible reality of type~I. Again, according to Fuchs \emph{et al.}\ \cite{FMS},
\bq
Instruments are the Copenhagen surrogate for experience. Being objective and independent of the agent using them, instruments miss the central point of QBism, giving rise to the notorious measurement problem, which has vexed physicists to this day.
\eq 
In actual fact, it is QBism that misses the central point of instruments. The possibility of drawing a line of separation between subject and object, in such a way as to be able to refer to objects without referring to experiences, is a \emph{sine qua non} of empirical science. This possibility calls for a language that is suitable for the unambiguous communication of ``what we have done and what we have learned'' (Refs.\ \cite[ pp.\ 39, 72, 89]{Bohr-APHK} and \cite[pp.\ 3, 24]{Bohr-EAPHK}).  The need for this language has nothing to do with the particular field of experience to which it is applied. We especially need to use it when we stumble upon a new field of experience,  so that ``there is, strictly speaking, no new observational problem in atomic physics'' \cite[p.\ 89]{Bohr-APHK}. There is no new observational problem because in quantum physics we are doing what we have always done: setting up experiments and reporting their results, in suitable language, which Bohr used to call ``plain language,'' ``classical language,'' or ``the language of classical physics.'' (Bohr never asserted that quantum mechanics requires classical physics itself.)

What \emph{is} new, and radically so, is that the properties of quantum systems are defined by the experimental conditions in which they are measured---i.e., by instruments---and that they only exist if and when they are measured. (Measurements are not confined to physics laboratories. Any object capable of providing information about a quantum system functions as a measuring device.) Quantum mechanics opens up a new and wholly unanticipated field of experience, revealing a quantum reality through statistical correlations between (classically describable) preparations and (classically describable) outcomes---i.e., between the properties and readings of instruments. It is these correlations that the mathematical apparatus of quantum mechanics serves to calculate: ``the physical content of quantum mechanics is exhausted by its power to formulate statistical laws governing observations obtained under conditions specified in plain language''~\cite[p.~12]{Bohr-EAPHK}.  

As a calculus of correlations, quantum mechanics presupposes the events to which, and on the basis of which, it serves to assign probabilities. It thus presupposes the existence of objects with outcome-indicating properties, and it rules out the application of the superposition principle to different outcome-indicating properties since it is \emph{logically inconsistent}, it is \emph{self-contradictory}, for an outcome-indicating device to  enter a superposition of different outcome-indicating properties (because then it would not be an outcome-indicating device). The standard objection to this point is that quantum mechanics has no way of telling us when an observable serves to indicate an outcome, and hence when it is capable of entering a superposition of its eigenstates. Therefore every observable must ``in principle'' be capable of entering such a superposition. Yet if quantum mechanics is a calculus of correlations, then it presupposes both the events to which and the events on the basis of which it serves to assign probabilities. It presupposes observables that are \emph{not} capable of entering superpositions of their eigenstates. If such observables did not exist, quantum mechanics would be a subject without a subject matter. 

It is tempting to attribute the exemption of outcome-indicating observables from the superposition principle to the intrinsic definiteness of our experience, which underlies Kant's principle of thoroughgoing determination. QBism is the latest and most uncompromising version of this strategy. Yet to subject everything but an individual's ``own direct internal awareness of her own private experience'' to the superposition principle, as QBists do, is overkill. What QBists rightly object to is the \emph{duplication} of instruments that results from conceiving them once as ``objective and independent of the agent using them'' and once again as aspects of our experience. It won't do to take refuge in the latter way of thinking about instruments only when the former (which allows outcome-indicating observables to enter superpositions of their eigenstates) leads to absurdity or poses a certifiably insoluble problem. According to the QBist, the instruments we experience are the only instruments there are. 

It would, however, be more correct to say that the instruments we experience are the only instruments \emph{we can have knowledge of and talk about}. But if (and insofar as) we are in possession of communicable knowledge of instruments, instruments are objective, not in the sense of belonging to the empirically inaccessible type-I reality but in the sense of forming part of the empirical type-II reality. Niels and Albert can talk about an apparatus without distinguishing between the apparatus experienced by Niels and the apparatus experienced by Albert and thus without reference to their respective experiences. One begins to understand the truth behind the rather absurd claim by Mermin (``a QBist in the making''~\cite{QBistinMaking}) that ``All versions of Copenhagen objectify each of the diverse family of users of science into a single common piece of apparatus''~\cite{Mermin_QBism}. 

Thus even though outcome-indicating properties are exempt from the superposition principle because they are experienced, they are nevertheless objective, albeit in the empirical rather than the transcendental sense. While they are independent neither of human experience nor of the community of users of science, they are independent of the specific experiences of any particular user.

\section{The QBist Agenda}\label{sec_QBism}
Three points dominate the QBist agenda: the construal of quantum-mecha\-nical probabilities as single-user Bayesian probabilities, the intention to put ``the scientist back into science'' so as to restore ``the balance between subject and object''~\cite{Mermin_Nature}, and the demonstration that quantum mechanics is ``\emph{explicitly} local''~\cite{FMS}. 

The Bayesian understanding of probability as a degree of confidence or belief denies the existence of external criteria---external to the \emph{individual}---for declaring a probability judgment right or wrong. The only criterion is coherence between the individual's beliefs. Because, in fact, there are such criteria---to wit, the objective measurement outcomes on the basis of which probabilities are assigned---the Bayesian construal of probability is unsuited for quantum mechanics. It is true that different individuals may assign probabilities on the basis of different objective outcomes and, hence, may use different quantum states, but this does not affect the objective nature of the events on which their probability assignments are based. These events are external to the \emph{individual} without being external to the community of users of science (i.e., without belonging to the transcendental reality of type~I.)

Again, the need to put ``the scientist back into science'' arises only if one assumes that science is concerned with a mind-independent (type-I) reality, as Schr\"odinger did in his essay ``Nature and the Greeks''~\cite[pp.\ 95--97]{Schroedinger-Nature-Greeks}, in which he deplored that ``I actually do cut out my mind when I construct the real world around me'':
\bq
the scientific picture of the real world around me is very deficient. It \dots\ is ghastly silent about all and sundry that is really near to our heart, that really matters to us. It cannot tell us a word about red and blue, bitter and sweet, physical pain and physical delight; it knows nothing of beautiful and ugly, good or bad, God and eternity\dots. And the reason for this disconcerting situation is just this, that, for the purpose of constructing the picture of the external world, we have used the greatly simplifying device or cutting our own personality out, removing it.
\eq
For Fuchs \emph{et al.}~\cite{FMS} this is reason enough to claim that Schr\"odinger ``takes a QBist view of science.'' In reality, though, Schr\"odinger merely objects to the transcendental objectivist view of science. If one takes science to be concerned with the empirical (type-II) reality objectified by us, there is nothing to be deplored, for the fact that the objectifying subjects, their quality-rich experiences, and their ethical and aesthetic values cannot be objectified, does not mean that they are therefore unreal.

It is worth mentioning that the qualitative aspects of our experience include the qualitative aspects of time and space. Weyl~\cite{Weyl1922} made this point with respect to space, stressing ``with what little right mathematics may claim to expose the intuitional nature of space'': 
\bq
Geometry contains no trace of that which makes the space of intuition what it is in virtue of its own entirely distinctive qualities which are not shared by ``states of addition-machines'' and ``gas-mixtures'' and ``systems of solutions of linear equations.'' It is left to metaphysics to make this ``comprehensible'' or indeed to show why and in what sense it is incomprehensible. We as mathematicians \dots\ must recognise with humility that our conceptual theories enable us to grasp only one aspect of the nature of space, that which, moreover, is most formal and superficial.
\eq
Much the same applies to time. What we can objectify is the spatial relations between objects and the temporal relations between states or the spatiotemporal relations between events. What we cannot objectify is the qualitative character of the warp and woof of our experience, phenomenal space and phenomenal time, including our subjective \emph{here} and \emph{now}. As I cannot objectify the particular point in space from which I survey my surroundings---a fact that nobody seems to deplore---so I cannot objectify my \emph{now} as a particular point in time at which I remember a past and anticipate a future---a fact that Einstein famously regretted~\cite{Carnap} and that Mermin~\cite{Mermin_Nature,Mermin2013} mistakenly believes QBism can redress.%
\footnote{There remains the question why a QBist should be worried about this. What does it matter if the qualitative aspects of our experience are not objectifiable, or not quite as objectifiable as the the quantitative ones? Are they therefore unreal? Only a greedy reductionist or a transcendental objectivist would think so.} 

The third point on the QBist agenda is to demonstrate that quantum mechanics is ``\emph{explicitly} local''~\cite{FMS}. The demonstration offered consists in the claim that ``space-like separated events \dots\ cannot be experienced by any single agent,'' and that therefore quantum correlations ``refer only to time-like separated events: the acquisition of experiences by any single agent.'' By giving ``each quantum state a home \dots\ localized in space and time---namely, the physical site of the agent who assigns it,'' QBism ``expels once and for all the fear that quantum mechanics leads to `spooky action at a distance'.''~\cite{Fuchs-QBatP}

In point of fact, by situating each quantum state at the physical site of the agent who assigns it, QBism contradicts its claim that, with the exception of the agent's ``own direct internal awareness of her own private experience''~\cite{FMS}, all external systems, including other agents, must be treated quantum-mechanically. If the agent has an objective position (a physical site at which she is located), then the pointer needle that serves to indicate an outcome, too, has one. If the needle has a position only in the mind of the agent who experiences it, then the agent can have a position only in the mind of someone who experiences her. One cannot have it both ways.

It is not that easy to expel the fear that quantum mechanics leads to spooky action at a distance. Suppose that Alice and Bob, having performed a large number of spin measurements on particles prepared ``in'' the singlet state, get together to compare notes.  What they find is clear evidence of the correlations that are predicted by the singlet state. The claim that quantum correlations refer only to time-like separated events is therefore wrong. The ``protection'' provided by ``truly personal quantum-state assignments'' \cite{Fuchs_Perimeter} does nothing to prevent the specter of action at a distance from being there ``as doggedly as it ever was.'' The reason local explanations---whether in terms of a common cause or via action at a distance---do not work needs to be understood, not swept under the rug by declaring that quantum correlations ``by their very nature'' \cite{FMS} only refer to time-like separated events.

\section{Semantic consistency}\label{Sec_sc}
\bq
\emph{The measurement problem is a \emph{consistency problem}. What we have to show is that the dynamics, which generally produces entanglement between two coupled systems, is consistent with the assumption that something definite or determinate happens in a measurement process. The basic question is whether it is consistent with the unitary dynamics to take the macroscopic measurement ``pointer'' or, in general, the macroworld as definite.}\par
{\parskip=0pt \hfill--- Jeffrey Bub~\cite{Bub}}
\eq
Can the semantic consistency of the theory be demonstrated without implicitly referring to the intrinsic definiteness of our experience? The aim of this section is to demonstrate that it can. The term ``semantic consistency'' was coined by von Weizs\"acker \cite[p.\ 260]{vWstr}. By the semantic consistency of a theory he meant ``that its preconceptions, how we interpret the mathematical structure physically, will themselves obey the laws of the theory.'' Thus if quantum mechanics is semantically consistent, even outcome-indicating observables will obey the theory's correlation laws---even though they are exempt from the superposition principle, and this not only FAPP (``for all practical purposes'').

Bub~\cite{Bub} has claimed that  ``unitary quantum dynamics'' can be made consistent with the existence of measurement outcomes if we reject the eigenvalue-eigenstate link and in its place stipulate that ``the decoherence `pointer' selected by environmental decoherence'' is always definite. Decoherence then ``guarantees the continued definiteness or persistent objectivity of the macroworld.'' Decoherence, however, merely displaces the coherence of the system composed of apparatus and object system into the degrees of freedom of the environment, causing the objectification problem to reappear as a statement about the system composed of environment, apparatus, and object system. Because the mixture obtained by tracing out the environment does not admit an ignorance interpretation, it can resolve the problem only FAPP.

In order to demonstrate the semantic consistency of quantum mechanics beyond ``FAPP,'' we must desist from conceiving of the objective world as if it were accessible to our senses \emph{on all scales}, and therefore spatially differentiated ``all the way down,'' which is what formulations of the theory that feature deterministically evolving quantum states implicitly assumes. The demonstration begins by showing that quantum mechanics itself \emph{implies} the incompleteness of the world's spatiotemporal differentiation. 

While quantum mechanics can tell us that the probability of finding a particle in a given region of space is~1, it is incapable of  giving us a region of space. For this a detector is needed. A detector is needed not only to indicate the presence of a particle in a region but also---and in the first place---to physically realize a region, so as to make it possible to attribute to a particle the property of being inside. Speaking more generally, a macroscopic apparatus is needed not only to indicate the possession of a property by a quantum system but also---and in the first place---to make a set of properties available for attribution to the system. (All of this is vintage Bohr.)

But if detectors are needed to realize regions of space, space cannot be intrinsically partitioned. If we conceive of it as partitioned, we can do so only as far as regions of space can be realized---i.e., to the extent that the requisite detectors are physically possible. This extent is limited by the indeterminacy principle, inasmuch as this rules out the existence of detectors with arbitrarily small sensitive regions that are (and remain) sharply localized relative to each other. If such regions cannot be realized (as the sensitive regions of detectors) then they are not available for attribution (as positions). Hence, if conceptually we keep partitioning space into smaller and smaller regions, we will reach a point beyond which the distinctions we make between regions no longer correspond to anything in the physical world. We can conceive of a partition of the physical world into \emph{finite} regions so small that none of them can be attributed (as a position) because none of them is available for attribution. In other words, physical space cannot be realistically modeled as an actually existing manifold of intrinsically distinct points. In yet other words, the spatial differentiation of the physical world is incomplete---it does not go ``all the way down.'' 

The same goes for the world's temporal differentiation, and this not only because of the relativistic interdependence of distances and durations. Just as the properties of quantum systems or the values of quantum observables need to be realized---made available for attribution---by macroscopic devices, so the times at which properties or values are possessed need to be realized by macroscopic clocks. And just as it is impossible for macroscopic devices to realize sharp positions, so it is impossible for macroscopic clocks to realize sharp times~\cite{Hilgevoord98}. Hence, neither the spatial nor the temporal differentiation of the physical world goes ``all the way down.''

In a world that is incompletely differentiated spacewise, the next best thing to a sharp trajectory is a trajectory that is so sharp that the bundle of sharp trajectories over which it is statistically distributed is never probed. In other words, the next best thing to an object with a sharp position is an object whose position probability distribution is and remains so narrow that there are no detectors with narrower position probability distributions---detectors that could probe the region over which the object's position extends. If the spatiotemporal differentiation of the physical world does not go ``all the way down,'' such objects must exist.%
\footnote{While decoherence arguments can solve the objectification problem only FAPP, they quantitatively support the existence of such objects.}
If I call them ``macroscopic objects,'' and if I call their positions ``macroscopic positions,'' it is not intended to mean that they are so large and/or massive as to behave like classical objects FAPP, but in the more rigorous sense just spelled out. By the ``macroworld'' I shall mean the totality of macroscopic positions.

What can be deduced from this characterization of macroscopic positions is that the events by which their values are indicated are (diachronically) correlated in ways that are consistent with the laws of motion that quantum mechanics yields in the classical limit. For any given time $t$ the following holds: if every event that indicates a macroscopic position prior to the time $t$ were taken into account, then---given the necessarily finite accuracy of position-indicating events---every event that indicates a macroscopic position at a later time would be consistent with all earlier position-indicating events and the classical laws of motion. There is, of course, one necessary exception: in order to permit a macroscopic pointer to indicate the value of an observable, its position must be allowed to change unpredictably if and when it serves to indicate an outcome.

Macroscopic objects thus follow trajectories that are only counterfactually indefinite. Their positions are ``smeared out'' only in relation to an imaginary spatiotemporal background that is more differentiated than the physical world. No value-indicating event reveals the indefiniteness of a macroscopic position in the only way it could, through a departure from what the classical laws predict. The testable correlations between outcomes of measurements of macroscopic positions are therefore consistent with \emph{both} the classical and the quantum laws \cite{ujm-opqf, Mohrhoff_manifesting}, and this makes it unnecessary to invoke the intrinsic definiteness of our experience in order to account for the existence of measurement outcomes.

\section{In search of  ``quantum reality''}\label{sec_qr}
\bq
\emph{I received a telephone call one day at the graduate college at Princeton from Professor Wheeler, in which he said, ``Feynman, I know why all electrons have the same charge and the same mass.'' ``Why?'' ``Because, they are all the same electron!''}\par
{\parskip=0pt \hfill--- Richard P. Feynman~\cite{Feynman_Nobel}}
\eq
We are now in a position to think about the objectifiable implications of the quantum-mechanical correlation laws without having to keep in mind our role as experiencing and objectifying subjects. The safest way to proceed is to use that formulation of quantum mechanics which carries the least metaphysical baggage, which is due to Feynman \cite{FHS}. It carries the least metaphysics baggage because all it essentially does is assign probability amplitudes to alternatives, which are defined as sequences of measurement outcomes, or as the continuum limits of such sequences. 

Both the wave-function formulation and Feynman's feature a pair of dynamical principles. In the former they are unitary evolution and collapse/objectification. In the latter they are summation over amplitudes (followed by taking the absolute square of the sum) and summation over probabilities (preceded by taking the absolute square of each amplitude). From the wave-function point of view, unitary evolution seems natural; what calls for explanation is collapse (the projection postulate) and objectification (postulated via the eigenvalue--eigenstate link). From Feynman's point of view, adding probabilities seems natural, inasmuch as this conforms to classical probability theory; what calls for explanation is why we have to add amplitudes. What is at issue, therefore, is not what causes wave functions to collapse and measurement to have outcomes but why we have to add amplitudes whenever quantum mechanics requires us to do so. In answer to this question I have proposed the following interpretive principle: 
\begin{itemize}
\item[(I)] Whenever quantum mechanics requires us to add amplitudes, the distinctions we make between the alternatives correspond to nothing in the physical world \cite{Mohrhoff_manifesting, Mohrhoff-QMinnewlight}.
\end{itemize}
This is a statement about the structure or constitution of the physical world, not a statement merely of our practical or conceptual limitations. Thus while the wave-function formulation stumps us with the dual problem of collapse and objectification, Feynman's formulation presents us with a question to which there is a clear answer. The reason why quantum mechanics requires us to add amplitudes is that, whenever it does, the distinctions we make between alternatives cannot be objectified (represented as real). The issue is not how measurement outcomes become objective but why distinctions we tend to make \emph{cannot} be considered objective.

We shall apply the  new interpretive principle to two paradigmatic set\-ups, one concerning distinctions between \emph{regions of space}, the other concerning distinctions between \emph{things}. In the context of a two-slit experiment (or any two-way interferometer experiment, for that matter), (I)~tells us that the distinction we make between $p_L$ = ``the particle went through the left slit ($L$)'' and $p_R$ = ``the particle went through the right slit ($R$)'' corresponds to nothing in the physical world. In \emph{some} sense the particle took both slits, but not in the sense of the conjunction $p_L\wedge p_R$, for one never sees a particle emerging both from the left slit and from the right slit. To say that the particle went through both slits can only mean that it went through the union of the two slits without going through a particular slit and without being divided into parts that go through different slits. This confirms our earlier conclusion that physical space cannot be an intrinsically differentiated expanse; if the parts of space defined by the slit were \emph{intrinsically} distinct, a particle could not go through both slits in the sense just spelled out.%
\footnote{For a more detailed argument see \cite[Sec.~4]{Mohrhoff_manifesting}.}

Now consider the elastic scattering of two identical particles initially moving northward and southward, respectively. The probability of eventually finding one particle moving eastward and one particle moving westward has the form
$$
\bigl|\braket{EW}{NS}\pm\braket{WE}{NS}\bigr|^2,
$$
where the sign depends on whether the particles are bosons or fermions. Because amplitudes are added rather than probabilities, (I) applies, and it tells us that the distinction we make between the alternative identifications

\medskip\centerline{$N=E,S=W$\quad or\quad $N=W,S=E$}

\medskip\noindent corresponds to nothing in the physical world. There is no answer to the question: ``Which outgoing particle is identical with which incoming particle?'' What does this tell us about the reality revealed by the quantum-mechanical correlation laws? The answer I propose is that the incoming particles (and therefore the outgoing ones as well) are \emph{one and the same entity}. Initially we observe two distinct properties, the property of moving northward and the property of moving southward, and subsequently we again observe two distinct properties, the property of moving eastward and the property of moving westward. We do not really observe two distinct things instantiating two distinct properties---this is one ``two'' too many. What we observe is perfectly consistent with saying that, both initially and subsequently, we observe the same thing twice, with distinct properties.%
\footnote{If this seems counterintuitive, it is because we tend to think of the ``parts of space'' as self-existent (rather than as being distinct regions only by virtue of being separately realized), and that we tend to think of whatever is contained in distinct regions as separately and independently existing things: ``things claim an existence independent of one another'' whenever they ``lie in different parts of space'' \cite{Einstein48}.}
What's more, there is no compelling reason to believe that this identity ceases when it ceases to have observable consequences owing to the presence of individuating properties. We are free to take the view that \emph{intrinsically} each particle is numerically identical with every other particle. What presents itself here and now with these properties and what presents itself there and then with those properties is one and the same entity. In what follows I shall call it ``Being.'' If you prefer any other name, be my guest.%
\footnote{According to French~\cite{SEP-French}, quantum mechanics is ``compatible with two distinct metaphysical `packages,' one in which the particles are regarded as individuals and one in which they are not.'' Esfeld~\cite{Esfeld2013} disagrees: it is not ``a serious option to regard quantum objects as possessing a primitive thisness (haecceity) so that permuting these objects amounts to a real difference.''}

\section{Manifestation}\label{sec_manif}
Because it is necessary to distinguish between (i)~properties that only exist if and when they are measured and (ii)~properties that are capable of indicating outcomes, the distinction between a classical or macroscopic domain and a non-classical or quantum domain is amply justified. But how can we understand the relation between the two domains, beyond the linguistic necessity of speaking about the quantum domain in terms of correlations between macroscopic events, which was stressed by Bohr \cite{Bohr-ATDN}? The answer I propose is that the distinction between the two domains is essentially a distinction between the manifested world and its manifestation.  

One of the reasons it is so hard to make sense of the quantum theory is that it answers a question we are not in the habit of asking. Instead of asking what the ultimate constituents of matter are and how they interact and combine, we need to broaden our repertoire of explanatory concepts and inquire into the \emph{manifestation} of the familiar world of everyday experience. Since the kinematical properties of microphysical objects---their positions, momenta, energies, etc.---only exist if and when they are indicated by the behavior of macroscopic objects, microphysical objects cannot play the role of constituent parts. They can only play an instrumental role in the \emph{manifestation} of macroscopic objects. 

Because the manifestation of the macroworld includes the manifestation of both space and time, it cannot be conceived as a process that takes place in space and time. We keep looking for the origin of the universe at the beginning of time, but this is an error of perspective. I propose to identify the origin of the universe with the Being that was introduced in the previous section, a Being intrinsically undifferentiated, transcendent of spatial and temporal distinctions. By entering into reflexive spatial relations---i.e., self-relations, relations between numerically identical relata---this Being gives rise to
\ben
\item what looks like a multiplicity of relata if the reflexive quality of the relations is ignored,%
\footnote{Because the relations are reflexive, the multiplicity of the relata is apparent rather than real. Does this mean that the material world is unreal, as some illusionist philosophies assert? By no means, for the material world owes its existence to a multitude of reflexive relations, and these are real.}
and 
\item what looks like a substantial expanse if the spatial quality of the relations is reified.
\een
As Leibniz said, \emph{omnibus ex nihilo ducendis sufficit unum}---one is enough to create everything from nothing.

The view put forward here goes farther in relationism---the doctrine that space and time are a family of spatial and temporal relations holding among the material constituents of the universe---in that it also affirms that the ultimate material constituents are numerically identical and formless. While fundamental particles are routinely described as pointlike, what is meant is that they lack internal structure. Lack of internal structure is suggested by the scale-invariance of a particle's effective cross-section(s) in scattering experiments with probe particles that are themselves pointlike in this sense, but it can be verified only down to the de Broglie wavelength of the probe particles. Hence, there can be no evidence of absence of internal structure, let alone evidence of a literally pointlike form. For further reasons why fundamental particles ought to be conceived as formless, see \cite[Sect.~9]{Mohrhoff_manifesting}. Conceived accordingly, the shapes of things resolve themselves into sets of spatial relations between formless and numerically identical relata. The truism that the universe lacks a position because it lacks \emph{external} spatial relations thus has a fitting complement: a fundamental particle lacks a form because it lacks \emph{internal} spatial relations.

The manifestation of the familiar world of everyday experience consists in a transition from the undifferentiated state of Being to a state that allows itself to be described in the classical language of interacting objects and causally related events, a transition from absolute unity to the multiplicity of the macroworld. Seen in this light, quantum theory reverses the explanatory arrow of common sense and classical or folk physics: instead of explaining wholes in terms of interacting parts, it suggests that the multiplicity of the world emerges from an intrinsically undifferentiated reality.

The transition from the absolute unity of Being to the multiplicity of the macroworld passes through several stages. Across these stages, the world's differentiation into distinguishable regions of space and distinguishable objects with definite properties is progressively realized. There is a stage at which Being presents itself as a multitude of formless particles. This stage is probed by high-energy physics and known to us through correlations between the counterfactual clicks of imagined detectors, i.e., in terms of transition probabilities between in-states and out-states. There are stages that mark the emergence of form, albeit as a type of form that cannot yet be visualized. The forms of nucleons, nuclei, and atoms can only be mathematically described, as probability distributions over abstract spaces of increasingly higher dimensions. At energies low enough for atoms to be stable, it becomes possible to conceive of objects with fixed numbers of components, and these we describe in terms of correlations between the possible outcomes of unperformed measurements. The next stage---closest to the manifested world---contains the first objects with forms that can be visualized---the atomic configurations of molecules. But it is only the final stage---the manifested macroscopic world---that contains the actual detector clicks and the actual measurement outcomes that have made it possible to discover and study the correlations that govern the quantum domain.

One begins to understand why the general theoretical framework of contemporary physics is a probability calculus, and why the probabilities are assigned to measurement outcomes. If quantum mechanics concerns a transition through which the differentiation of the world into distinguishable objects and distinguishable regions of space is gradually realized, the question arises as to how the intermediate stages are to be described---the stages at which the differentiation is incomplete and the distinguishability between objects or regions of space is only partially realized. The answer is that whatever is not completely distinguishable can only be described by assigning probabilities to what is completely distinguishable, namely to the different possible outcomes of a measurement. What is instrumental in the manifestation of the world can only be described in terms of (correlations between) events that happen or could happen in the manifested world.

Quantum mechanics thus presents us with a so far unrecognized kind of causality,%
\footnote{Unrecognized, I believe, within the scientific literature albeit well-known to metaphysics, inasmuch as the general philosophical pattern of a single world-essence (``Being'') manifesting itself as a multiplicity of individual things is found throughout the world. Some of its representatives in the Western hemisphere are the Neoplatonists, John Scottus Eriugena, and the German idealists. The quintessential Eastern example is the original (pre-illusionist) Vedanta of the Upanishads \cite{Phillips, SA_Isha, SA_Kena}.}
which must be distinguished from its more familiar temporal cousin.%
\footnote{While an atemporal causality does not involve a temporal sequence, process, or transition, it does involve stages by which the differentiation of the world into distinguishable objects and distinguishable regions is progressively realized.}
The usefulness of the latter, which links states or events across time or spacetime, is confined to the world drama; it plays no part in setting the stage for it. It helps us make sense of the manifested world as well as of the cognate world of classical physics, but it throws no light on the process of manifestation nor on the quantum correlations that are instrumental in the process.

The causality associated with the atemporal process of manifestation---the transition from the undifferentiated state of Being to the multiplicity of the macroworld---presents the nonlocality of quantum mechanics in a new light. The atemporal process by which Being enters into reflexive relations and matter and space come into being as a result, is the nonlocal event \emph{par excellence}. Depending on one's point of view, it is either coextensive with spacetime (i.e., completely delocalized) or ``outside'' of spacetime (i.e., not localized at all). Occurring in an anterior relation to space and time, it is the common cause of all correlations, not only of the seemingly inexplicable ones between simultaneous events in different locations but also of the seemingly explicable ones between successive events in the same location.%
\footnote{It seems to me that the diachronic correlations between events in timelike relation are as spooky as the synchronic correlations between events in spacelike relation. While we know how to calculate either kind, we know as little of a physical process by which an event here and now contributes to determine the probability of a \emph{later} event \emph{here} as we know of a physical process by which an event here and now contributes to determine the probability of a \emph{distant} event \emph{now}.}

When I posit an intrinsically undifferentiated (and hence unqualifiable) Being, which manifests the world by entering into reflexive spatial relations, am I not referring to a transcendental (type-I) reality rather than an empirical (type-II) reality? The answer is: not intentionally and not knowingly. As the Greek philosopher--poet Xenophanes pointed out some twenty-five centuries ago, even if our conceptions represented the world exactly as it is, we could never know that this was the case. What can be said is this: the fact that our models of reality are mental constructs abstracted from our experiences does nothing to explain why most of our theoretical constructs turn out to be non-objectifiable. There has to be something about the (type-I) reality at the origin of our our selves and our experiences that licenses the objectification of only a very limited set of mental constructs. Although a constructed empirical reality can never \emph{match} the transcendental reality, the former may \emph{fit} the latter.

The distinction between a match and a fit, which is due to von Glasersfeld \cite{vG}, has been illustrated by the allegory of a skipper who, in the dark of a stormy night, without navigational aids, passes a narrow strait whose contour he does not know \cite{Watzlawick}. Epistemologically we are in the skipper's position. If he reaches the open sea without mishap, he has found a course that \emph{fits} the strait; if next time he takes the same course, he will again pass safely. What he has not obtained is a map that \emph{matches} the coastline. To precisely locate at least one point of the coastline, he must come into contact with it---at the risk of wrecking his ship. My contention is that the ideas put forward here and in greater detail in Refs.\ \cite{Mohrhoff_manifesting, Mohrhoff-QMinnewlight} \emph{fit} the reality at the origin of our experiences. No scientific theory can achieve more than that.

\section{Experience}\label{sec_exp}
\bq
\emph{Nobody has the slightest idea how anything material could be conscious. Nobody even knows what it would be like to have the slightest idea about how anything material could be conscious.}\par
{\parskip=0pt \hfill--- Jerry A. Fodor~\cite{Fodor}}
\eq
Experience being the matrix of empirical knowledge, it cannot become an object of empirical knowledge. If we seize upon quantitative aspects of our experience and think of them as forming a mind-independent reality, it should not perplex us that this reality cannot contain or give rise to the quality-rich experience we have discarded in the process. 

The most promising alternative to reductionism in the philosophy of mind is panpsychism \cite{Griffin, Sprigge, Rosenberg, Skrbina, Strawson}. Beginning with Leibniz in the 17th Century, philosophers have argued that all physical properties are relational or extrinsic, and none are in a fundamental sense non-relational or intrinsic. This offers the possibility of situating consciousness among the intrinsic properties of the relata which bear the relational properties. This possibility has been considered by Bertrand Russell~\cite{Russell} and more recently by Chalmers~\cite{Chalmers}. The problem with it is that it is hard to imagine how the consciousnesses of a myriad of particles can constitute the unified consciousness that we enjoy. If however we take into account not only that all physical properties are relational but also that all relational properties are reflexive, so that the relata are identically the same Being, the concept of  consciousness as an intrinsic aspect of the relata comes into its own.

It stands to reason---at any rate, it makes good sense---that Being does not simply manifest the world; rather, Being manifests the world \emph{to itself}. Being relates to the world not only as the substance that constitutes it but also as the consciousness that contains it. It is at once the single substance by which the world exists and the ultimate self or subject for which it exists. How, then, might we, as conscious beings, be related to this ultimate self or subject? This question has been answered in considerable detail and on a solid experiential foundation by the Indian philosopher (and freedom fighter, and mystic) Sri Aurobindo \cite{Heehs2008}. In keeping with a more than millennium-long philosophical tradition \cite{Phillips}, Sri Aurobindo \cite{SA_LD} posits an Ultimate Reality whose intrinsic nature is (objectively speaking) infinite Quality and (subjectively speaking) infinite Delight. This has the power to manifest its inherent Quality/Delight in finite forms, and the closest description of this manifestation is that of an all-powerful consciousness creating its own content.

In the native poise of this consciousness, its single self is coextensive with its content and identical with the substance that constitutes the content. There, but only there, it is true that \emph{esse est percipi} (to be is to be perceived). A first self-modification of this \emph{supramental} consciousness leads to a poise in which the one and only self adopts a multitude of standpoints, localizing itself multiply within the content of its consciousness and viewing the same in perspective. It is in this secondary poise that the dimensions of experiential space (viewer-centered depth and lateral extent) come into being. It is also here that the dichotomy between subject and object, or self and substance, becomes a reality.%
\footnote{To assume that the ultimate object is one with the ultimate subject and to show how they become distinct may be the only way that the explanatory gap between object and subject \cite{Levine} can be bridged.}

Probably the most adequate description of the process by which the original self assumes a multitude of standpoints is that of a multiple concentration of consciousness. A further self-modification of the original creative consciousness occurs when this multiple concentration becomes exclusive. We all know the phenomenon of exclusive concentration, when consciousness is focused on a single object or task, while other goings-on are registered subconsciously, if at all. A similar phenomenon transforms individuals who are conscious of their essential identity into individuals who have lost sight of this identity and, as a consequence, have lost access to the supramental view of things. Their consciousness is mental, which means not only that it belongs to what appears to be a separate individual but also that it perceives or represents the world as a multitude of separate objects. Mentally conscious beings thus come into existence not only by an evolution from seemingly unconscious matter but also, and in the first place, by a multiple exclusive concentration of the creative consciousness inherent in Being.  \looseness=1

If this multiple exclusive concentration is carried to its logical conclusion, the result is a world whose inhabitants lack both the ability to generate ideas (which is a function of the principle of mind) and the power to execute them (which is a function of the principle of life).%
\footnote{A hundred years ago it seemed obvious to many that life could not have emerged from utterly lifeless matter, just as today it seems obvious to many that experience cannot have emerged from utterly non-experiential matter. Yet today no one appears to seriously doubt that life did emerge from utterly lifeless matter; the seemingly insuperable ``hard problem of life'' simply dissolved. So why should it not be the same with the ``hard problem of consciousness,'' a hundred years from now? As Strawson \cite{Strawson} has pointed out, one cannot draw such a parallel unless life is considered completely apart from conscious experience. If consciousness is essential to life, as may well be the case, one cannot reduce life to physics via chemistry if consciousness cannot be reduced along with it.}
And because the latter is also responsible for the existence of individual forms, the result is a world of formless individuals---the fundamental particles of physics. This is how the original creative consciousness came to be ``involved'' in mind, how mind came to be ``involved'' in life, and how life came to be ``involved'' in formless particles. And because these principles are ``involved'' in formless particles, matter is capable of evolving life, life is capable of evolving mind, and mental consciousness can and is bound to eventually evolve the supramental consciousness---the infinite conscious force by which Being manifests the world. 

The action of this supramental consciousness is primarily qualitative and infinite and only secondarily quantitative and finite. Essentially, mind is the agent of the supermind's secondary, quantifying, and delimiting action. But when it is separated in self-awareness from its supramental parent consciousness, as it is in us, it not only divides \emph{ad infinitum} but also takes the resulting multiplicity for the original truth or fact. This is why we tend to construct reality from the bottom up, on an intrinsically and completely differentiated space or space-time, out of locally instantiated physical properties, or else by aggregation, out of a multitude of individual substances. And that, at bottom, is why making sense of quantum mechanics is so hard, for (as we found in Sect.~\ref{Sec_sc}) the spatial differentiation of the physical world does not go ``all the way down.'' Reality is structured by a progressive self-differentiation of Being that does not ``bottom out,'' and this makes it impossible to construct reality from the bottom up.

\section{Concluding thoughts}
\bq
\emph{But the fact that we find ourselves in a quantum world where measurement is possible \dots\ will surely involve the same sort of explanation as the fact that we find ourselves in a world where we are able to exist as carbon-​​based life forms.}\par
{\parskip=0pt \hfill--- Jeffrey Bub~\cite[p.\ 234]{Bub-IQW}}
\eq
Quantum physics does not explain ``how nature does it.'' The theory only explains---via conservation laws---why certain things \emph{won't} happen. This is exactly what one would expect if the force at work in the world were an infinite (unlimited) force operating under self-imposed constraints. In that case we would have no reason to be surprised or dismayed by the impossibility of explaining the quantum-mechanical correlation laws in terms of physical mechanisms, for it would be self-contradictory to invoke a physical mechanism to explain the working of an infinite force. What would need explaining is why (for which purpose) this force works under the particular constraints that it does. Where efficient causality fails, final causality takes over. 

It stands to reason that setting the stage for the drama of evolution calls for the existence of sufficiently stable objects that have spatial extent---objects that ``occupy'' space. Because this stage has been set by carrying the multiple exclusive concentration of the consciousness one with Being to its logical conclusion, which resulted in a world of forms that resolve themselves into spatial relations between an apparent multitude of formless relata, such objects will appear to be ``made'' (i.e., manifested by means) of finite numbers of formless objects. But, as I have argued elsewhere \cite{Mohrhoff-justso, Mohrhoff-QMexplained, Mohrhoff-book},  the existence of objects that (i)~have spatial extent, (ii)~neither collapse nor explode as soon as they are formed, and (iii)~are composed of finite numbers of objects that lack spatial extent, not only requires the validity of quantum mechanics but also goes a long way toward establishing the other well-tested laws of contemporary physics.  The validity of quantum mechanics may thus be a necessary consequence of the manner in which---and the purpose for which---the physical world came into being.

In concluding, I would like to put in perspective our attempts to make sense of this particular world in which we find ourselves. We tend to think of the evolution of consciousness as the successive emergence of increasingly adequate ways of experiencing a world that exists \emph{in itself}, out of relation to any kind of consciousness or experience. And we tend to think that our way of experiencing this world is more or less adequate to the task of understanding it. But this is not likely to be the case. As Nagel \cite[p.~10]{Nagel} wrote, ``too many hypotheses and systems of thought in philosophy and elsewhere are based on the bizarre view that we, at this point in history, are in possession of the basic forms of understanding needed to comprehend absolutely anything.'' The mistake, however, does not lie in thinking that our ways of experiencing and understanding are adequate to the task of comprehending the world. The mistake is to think that there is a world which exists out of relation to any kind of consciousness or experience, in which consciousness can nevertheless emerge. There are only different ways in which Being presents itself to itself, including a Houdiniesque way in which its intrinsic consciousness and freedom are ``involved'' in a world that appears to be governed by mechanical necessity and chance.

Our very concepts of space, time, and matter are conditioned by and dependent on the manner in which we, at this point in history, experience the world---the manner in which Being presents itself to itself in the mode of experience to which it has attained in us. Matter, then, can not be the origin of consciousness.%
\footnote{The discussion by Hut \emph{et al.}~\cite{HAT} of the math-matter-mind triangle may be relevant in this context. The triangle suggests the circularity of the views that mathematics is a creation of mind, that mind arises out of matter, and that matter is governed by mathematical laws.}
It is consciousness that has created matter, first by carrying its multiple exclusive concentration to the point of being involved in a multitude of formless particles, and again by evolving our present mode of experiencing the world, which has given us the ability to integrate images into three-dimensional objects from which the experiencing subject can abstract itself---an ability that earlier modes of human consciousness did not possess~\cite{Gebser}. 

There are expressions of these earlier modes that reveal how they differ from our present mode. Consider, for instance, the ancient notion that the world is contained in a sphere, which has the fixed stars attached to its boundary, the firmament. We cannot but ask: what is outside that sphere? Those who held this notion could not, because for them the third dimension of perceptual space---viewer-centered depth---did not at all have the reality that it has for us. This is why they could not handle perspective in drawing and painting, and why they were unable to arrive at the subject-free stance which is a prerequisite of modern science. All this became possible only during the Renaissance.

While our present mode of consciousness has enabled us to discover much that is relevant to understanding our evolutionary past, our tendency to regard its experience of the world as the definite and final experience of what there is prevents us from envisioning modes of consciousness that transcend our present time- and space-bound mode. As the latter has enriched the pre-scientific mode with a new experiential dimension---viewer-centered depth---so will the next mode enrich our present mode, with the likely result that our present, scientific and philosophical understanding of the world will come to seem as dated as the mythological understanding of the pre-scientific era seems today. Just as the mythological understanding could not foresee the technological explosion made possible by science, so our scientific imagination is incapable of foreseeing the radical changes that might be wrought by the evolution of a new mode of experience.

%\nocite{*}

\end{document}